\documentclass[aps,prper,reprint,longbibliography,floatfix]{revtex4-2}   
\usepackage[T1]{fontenc}
\usepackage[utf8]{inputenc}
\usepackage{times}
\usepackage{array}
\usepackage{titlesec}

\usepackage{geometry}
\usepackage{array}
\newcolumntype{C}[1]{>{\centering\arraybackslash}m{#1}} 
\newcolumntype{L}[1]{>{\raggedright\arraybackslash}m{#1}} 
\newcolumntype{R}[1]{>{\raggedleft\arraybackslash}m{#1}} 

\usepackage{array}
\newcolumntype{P}[1]{>{\centering\arraybackslash}m{#1}}
\usepackage{booktabs}     
\usepackage{tabularx}     
\usepackage{ragged2e}     
\usepackage{caption}      
\usepackage{multirow}
\usepackage{placeins}
\usepackage{amsmath}
\usepackage{amssymb}
\usepackage{array} 
\newcolumntype{C}[1]{>{\centering\arraybackslash}m{#1}}  
\usepackage{graphicx}
\usepackage{tikz}

\usepackage{enumerate}
\usepackage{float}
\usepackage{array}
\usepackage{placeins}
\usepackage{hyperref}
\hypersetup{colorlinks=true,urlcolor=blue,citecolor=blue,linkcolor=blue}
\usepackage{comment}
\usepackage{indentfirst}
\usepackage{tcolorbox}
\usepackage[normalem]{ulem}  
\usepackage{setspace}        

\usepackage{float}
\usepackage{placeins}
\begin{document}

\begin{titlepage}

\title{Using an LLM to Investigate Students' Explanations on Conceptual Physics Questions}

 \author{Sean Savage}
 \affiliation{Department of Physics and Astronomy, Purdue University, 525 Northwestern Ave, West Lafayette, IN, 47907, U.S.A.}

 \author{N. Sanjay Rebello}
 \affiliation{Dept. of Physics and Astronomy / Dept. of Curriculum and Instruction, Purdue University, West Lafayette, IN, 47907, U.S.A.} 

\begin{abstract}
Analyzing students' written solutions to physics questions is a major area in PER.  However, gauging student understanding in college courses is bottlenecked by large class sizes, which limits assessments to a multiple-choice (MC) format for ease of grading. Although sufficient in quantifying scientifically correct conceptions, MC assessments do not uncover students' deeper ways of understanding physics. Large language models (LLMs) offer a promising approach for assessing students’ written responses at scale. Our study used an LLM, validated by human graders, to classify students' written explanations to three questions on the Energy and Momentum Conceptual Survey as correct or incorrect, and organized students' incorrect explanations into emergent categories. We found that the LLM (GPT-4o) can fairly assess students' explanations, comparable to human graders (0-3\% discrepancy). Furthermore, the categories of incorrect explanations were different from corresponding MC distractors, allowing for different and deeper conceptions to become accessible to educators. 

    \clearpage
  \end{abstract}

\maketitle
\end{titlepage}
\maketitle

\section{INTRODUCTION}
Research-based multiple choice (MC) conceptual assessments such as the Force Concept Inventory (FCI) \citep{hestenes1992force}, Force and Motion Conceptual Evaluation (FMCE) \citep{thornton1998assessing}, and the Energy and Momentum Conceptual Survey (EMCS) \citep{singh2016multiple} have long been used to assess students' conceptual understanding in physics, primarily for their multiple-choice format and ease of large-scale implementation \citep{nieswandt2009written}. 
While MC assessments are sufficient in gauging the percentage of students who have scientifically correct conceptions, they have several limitations on student learning \citep{Kuechler_Simkin_2003, Petersen2016, shepard2000role, Wong2021}\citep{Wood2021}. MC assessments are neither effective in identifying the myriad ways in which students think about the physics concepts \citep{rebello2004effect} nor the conceptual resources \citep{hammer2000student} that students activate in the context of questions in the inventory. Furthermore, prior studies have demonstrated that repeated students' exposure to distractors (incorrect MC options) strengthens incorrect conceptual associations\citep{Roediger2005}.

MC inventories offer a rich repertoire of questions designed to assess students' conceptual understanding of physics \citep{good2019}, which, if strategically employed, could provide insights into their underlying mental models \citep{bao2006model} and conceptual resources. Shifting from a fixed-choice options format to a written response assessment allows students to express their reasoning in their own words. Despite their pedagogical value, investigating students' written responses at scale, particularly in large-enrollment courses ($N > 1,000$), poses significant challenges. Achieving consistent and equitable grading of open-ended assessments across large student cohorts remains a prohibitive endeavor in the field. 

With increasing class sizes across universities, strategies are needed to assess students' written answers to physics questions' concepts. Recently, artificial intelligence (AI) has emerged as a promising tool for evaluating students' written responses due to its ability to analyze data and uncover patterns for large sets of data while being consistent in its analysis \citep{aleven2022ai,munsell2021using}. Furthermore, AI offers the potential to alleviate grader fatigue \citep{Casalino2021} while demonstrating effectiveness in examining student reasoning across several studies \citep{kortemeyer2023, usdoe2023ai, weijers2025intuition, Wan_Chen_2024}. Notably, organizations such as Khan Academy have recently begun to implement AI into their instructional tools, providing feedback to written responses for K-12 students \citep{khan2023}. Outside of PER, some studies \citep{BUTCHER2010489, Salim2022} have used computational methods to classify students' responses and have shown their potential to be equal to or greater than human graders. 

\begin{figure}[!htbp]
\fbox{\includegraphics[width=0.96\linewidth]{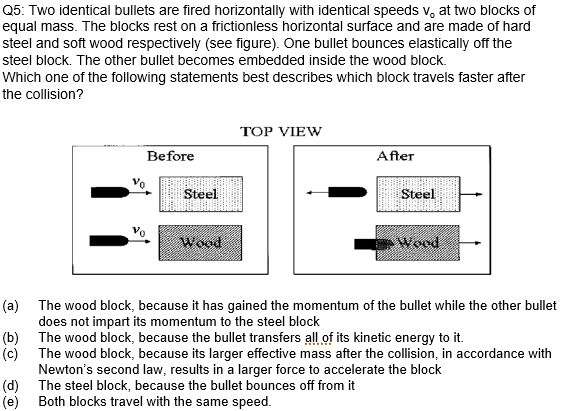}}
\caption{Question 5 from the EMCS (correct choice, D) \citep{singh2016multiple}.}
\label{fig:EMCS_Q5}
\end{figure}

\begin{figure}[!htbp]
\fbox{\includegraphics[width=0.96\linewidth]{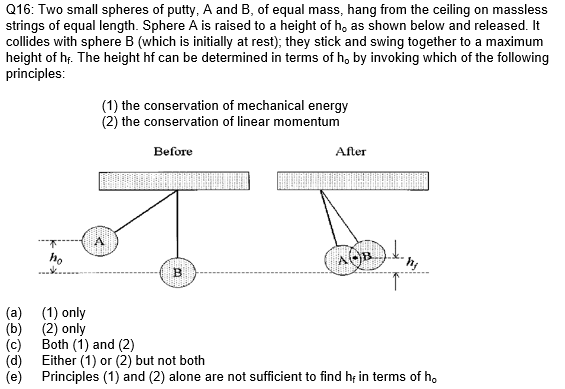}}
\caption{Question 16 from the EMCS (correct choice, C) \citep{singh2016multiple}.}
\label{fig:EMCS_Q16}
\end{figure}

\begin{figure}[!htbp]
\fbox{\includegraphics[width=0.96\linewidth]{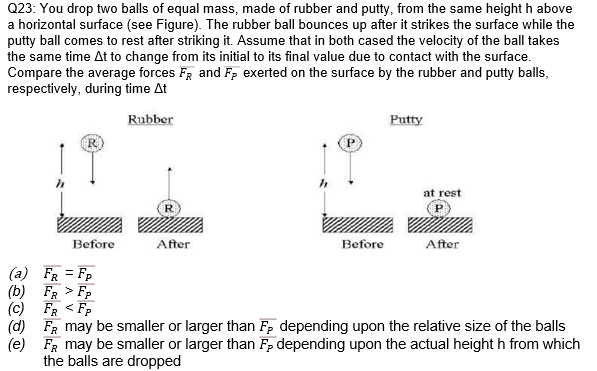}}
\caption{Question 23 from the EMCS (correct choice, B) \citep{singh2016multiple}.}
\label{fig:EMCS_Q23}
\end{figure}

This study aims to compare the methods for analyzing explanations that students wrote for conceptual physics questions on a widely used and well-established \citep{singh2016multiple} research-based assessment in introductory undergraduate physics – the Energy and Momentum Conceptual Survey (EMCS) \citep{singh2016multiple}. This study will use validated questions from EMCS, for which the consistency and strength of AI-based feedback will be gauged.  We will modify three MC questions (see Figs \ref{fig:EMCS_Q5},\ref{fig:EMCS_Q16},\ref{fig:EMCS_Q23}) from the EMCS into written questions for our analysis. Students will write their responses to the question using the claim, evidence, and reasoning framework \citep{mcneill2011supporting}, which was introduced to students in the class. Previous work has shown the strength of the EMCS as an assessment that is well constructed to accurately gauge the students’ understanding of energy and momentum concepts \citep{Afif2017}; therefore, it is an appropriate assessment to evaluate the claim (answer) along with evidence and reasoning (which together constitute and explanation for the answer).
\vspace{0.25em}

\vspace{-0.25em}
\vspace{-0.25em}

Requesting students to describe their claim, evidence, and reasoning will enable us to investigate the effectiveness of Large Language Models (LLMs) at gauging students' conceptual resources. Thus, our research questions are:

\vspace*{1em}
RQ1: \textit{How does the accuracy of a Large Language Model(LLM) in grading student essay responses to conceptual questions on the EMCS compare with human graders?} 

RQ2: \textit{What are the categories of students’ responses emergent from the analysis by an LLM of student essay responses to conceptual questions on the EMCS?}

\section{METHODS}
\subsection{Data Collection}
The present study was conducted in a first-semester calculus-based physics course at a large US Midwestern land-grant university. The EMCS was administered in the Spring semester as a pre-test in Week 1 of the course and a post-test in Week 12 after the class had covered the relevant concepts (momentum principle and energy principle). Both the pre- and post-tests were completed by students online, within a 60-minute time limit, and monitored using a virtual proctoring system \citep{brightspace}.  
Students were tasked to respond to the multiple-choice versions of all questions on the EMCS except for three questions -- Q05, Q16 and Q23 (Figs. \ref{fig:EMCS_Q5},\ref{fig:EMCS_Q16},\ref{fig:EMCS_Q23}) for which the original multiple-choice responses were removed and instead students were asked to provide a written response. These three questions (Q05, Q16, and Q23) were selected because they had shown some of the lowest pre- to post-test gains along with poor results in post test performance (27, 26, and 25\% respectively) in literature \citep{singh2016multiple}, meaning that students were most likely to activate unproductive conceptual resources to answer these questions even after learning the material. As such, our work exclusively utilized the post-test data collected from responses to all EMCS questions submitted by $N = 1131$ students (about 78\% of the enrollment).

Open-ended responses from students were collected in the form of a scientific argument, including their answer (i.e., a claim) as well as an explanation (i.e., evidence and reasoning) to support it. In the case of Q05, for example, students were instructed to: "\textit{Use claim, evidence, and reasoning to compare how fast the blocks are moving concerning each other after the collision. Use complete sentences}". Students were already familiar with constructing scientific arguments to answer conceptual questions as they had received considerable instruction on writing scientific arguments using the claim-evidence-reasoning framework \citep{mcneill2011supporting} in the Recitation component of the course since the beginning of the semester.

Notably, this study presented three questions (Q05, Q16, and Q23), out of 25, in a written format because requiring open-ended responses for all EMCS questions would be prohibitive in terms of the time needed to complete the survey and could lead to student fatigue.

\subsection{Data Analysis}
OpenAI's GPT-4o (one of the most up-to-date LLMs from OpenAI at the time \citep{openai2025chatgpt}) was called into a Jupyter Notebook to analyze the $N = 1131$ students' written responses. We prompted the GPT-4o  API (Application Programming Interface) to include criteria for classifying students as correct or incorrect based on their responses in terms of their claim and explanation (evidence/reasoning). Our prompt template was given the following:

\vspace*{0.5em}

\textit{\small
system = "You are an experienced physics educator grading a student's written response to a conceptual physics question. 
}

\vspace*{0.25em}

\textit{\small rubric= Before assigning scores, think through the student's response step by step. 
Explain how the student's claim and reasoning relate to the relevant physics principles.\\ -Claim: Score 1 if the student states or implies the correct outcome.\\
- Reasoning: Score 1 only if the evidence and reasoning appropriately use physics principles to support the claim.\\
- Overall: Score 1 only if the claim, evidence/reasoning are correct.\\
Assign reasoning-based categories that highlight distinct conceptual patterns or conceptions. Avoid lumping diverse conceptions into a broad category unless justified. Aim for up to 6 categories, including a category for “Correct Understanding”. "}

\vspace*{0.5em}

The code was built to take these prompts in, along with the specific question, into a prompt template which would then output the evaluation of students' essays. This analysis was carried out by running the code for each question at a time. As such, it was run continuously to keep the grading consistent. We set a temperature of 0.5 to have some leeway in creating new categories while still maintaining consistency.  This was reflected in the consistent emergent categories of students' responses and low standard deviation across runs in the percentage of students marked with fully correct scores (see Table \ref{Table_CorrectResponses}). The output of the code generated a .csv file containing, for each response, whether the student was correct in their claim (answer), in their evidence/reasoning (explanation), and overall correctness. It also categorized students' conceptual resources and gave an explanation of why they were unproductively applied. The prompt was constructed with role prompting ("...experienced physics educator...")  and chain of thought thinking ("Before assigning scores, think through the student's response step by step..."), both proven to be effective ways of prompting \citep{Chen2023}. In addition to the template above, the LLM prompt included instructions in the template such as: "Before assigning a category, identify which physics principle(s) the student applied (e.g., energy, momentum, force), evaluate whether they used all necessary principles..." Doing so takes advantage of the reasoning capabilities that the newer LLMs possess, as the text it will look at is based on a multi-step physics question.
In order to validate the accuracy of the LLM's decision on students' level of correctness for claim and evidence/reasoning, a human grader (HG) analyzed a randomly chosen 20\% ($N = 231$) sample of responses.

\section{FINDINGS \&\ DISCUSSION}
Table \ref{Table_CorrectResponses} shows the results for both the LLM's output and that of the human grader (HG). The discrepancy obtained between LLM and HG for the overall correctness i.e., correct claim (answer) supported by correct explanation (evidence and reasoning), was remarkably low (0-3\%). For the randomly chosen 20 \% ($N = 231$) responses, the mean Cohen's kappa was 0.77 for the correct claim/evidence/reasoning across the three questions ($\kappa = 0.74$, $0.65$, and $0.89$, respectively). These values are known to be substantial to almost perfect \citep{Sainani_2017}, reinforcing the strong agreement between LLM and HG. Interestingly, the LLM was less likely to mark a student as completely correct (Table \ref{Table_CorrectResponses}). 
\\

\begin{table}[htbp]
\centering
\captionsetup{justification=raggedright,singlelinecheck=false} 
\caption{\small Average percentage correctness rated by the LLM and HG and for overall correctness (claim/evidence/reasoning), along with standard deviation of LLM across 3 runs.}
\begin{ruledtabular}
\small
\begin{tabular}{lll}
\textbf{Q \#} & \textbf{LLM ($N = 1153$)} & \textbf{HG ($N = 231$)}\\
\hline
Q05  & $33 \pm 1.1$ \% & 34\% \\
Q16 & $27 \pm 1.6$ \% & 30\% \\
Q23 & $45 \pm 0.7$ \% & 45\%\\
\end{tabular}
\end{ruledtabular}
\label{Table_CorrectResponses}
\end{table}

In addition to the percentage of correct responses, we computed the distribution of incorrect responses as categorized by the LLM. These distributions are shown below for Q05 (Table \ref{Table_IncorrectExplanations_Q05}), Q16 (Table \ref{Table_IncorrectExplanations_Q16}), and Q23 (Table \ref{Table_IncorrectExplanations_Q23}). These make up the percentages of the $N = 1153$ students, with the remaining percentages coming from correct responses, Table \ref{Table_CorrectResponses}, and miscellaneous responses ranging from 1-2\%.

\begin{table}[htbp]
\centering
\captionsetup{justification=raggedright,singlelinecheck=false} 
\caption{\small Incorrect explanation categories on Q05.}
\begin{ruledtabular}
\small
\begin{tabular}{ll}
\textbf{Explanation} & \textbf{\% responses}\\
\hline
Misapplying Collision Type  & 34\% \\
Misapplying Momentum Principle & 26\% \\
Misapplying Work Energy Principle & 2\% \\
Misidentifying Collision Type  & 2\% \\
Misidentifying Newton's 2nd Law  & 1\% \\
\end{tabular}
\end{ruledtabular}
\label{Table_IncorrectExplanations_Q05}
\end{table}

\begin{table}[htbp]
\centering
\captionsetup{justification=raggedright,singlelinecheck=false} 
\caption{\small Incorrect explanations categories on Q16.}
\begin{ruledtabular}
\small
\begin{tabular}{ll}
\textbf{Explanation} & \textbf{\% responses}\\
\hline
Omitting Energy Conservation  & 37\% \\
Misapplying Energy Principle & 20\% \\
Omitting Momentum Conservation & 12\% \\
Misapplying Impulse Principle  & 2\% \\
Misapplying Momentum Principle  & 1\% \\
\end{tabular}
\end{ruledtabular}
\label{Table_IncorrectExplanations_Q16}
\end{table}

\begin{table}[htbp]
\centering
\captionsetup{justification=raggedright,singlelinecheck=false} 
\caption{\small Incorrect explanation categories on Q23.}
\begin{ruledtabular}
\small
\begin{tabular}{ll}
\textbf{Explanation} & \textbf{\% responses}\\
\hline
Misapplying Momentum Principle  & 17\% \\
Misapplying Collision Concepts & 13\% \\
Misapplying Force Concepts & 10\% \\
Misapplying Impulse Principle  & 7\% \\
Misapplying Energy Principle  & 6\% \\
\end{tabular}
\end{ruledtabular}
\label{Table_IncorrectExplanations_Q23}
\end{table}

\vspace{0.25em}
The API employed in this study was not instructed with sample predefined categories on common students' incorrect conceptions; rather, it was constrained to generate up to six emergent categories based solely on the input data. To validate the grading performance by the API, we prompted it with no category limit. Although niche situations appeared, the main categories persisted across multiple runs of the API for all questions, indicating a consistent grading performance. 

Notably, our analysis revealed that the response categories generated did not align with the predefined distractors in the MC format version. For instance, on Q05, the MC distractors include misapplying conservation of momentum (choice A), misapplying the conservation of energy (choice B), or misapplying Newton's Laws (choice C).  Some degree of overlap was observed between MC option A and the LLM categories of "\textit{Misapplying Momentum Principle}" and "\textit{Misidentifying Newton's 2nd Law}". However, several key LLM-derived categories for Question 5, such as '\textit{Misapplying Collision Type}' and '\textit{Misidentifying Collision Type}', were not reflected in the MC distractors (Table \ref{Example_Student_Response_Codes}). This strongly suggests that LLMs can reveal patterns of student reasoning that remain uncaptured by fixed-response options, underscoring the limitations of physics students' conceptions within the limits of MC formats.

For questions Q16 and Q23, the traditional MC format proved particularly limiting in eliciting insights into students' conceptual understanding, as the answer choices were restricted to identifying which principles were applicable. While the LLM-generated categories did capture omissions of relevant principles, they also revealed instances where principles were incorrectly applied, an equally important dimension of student reasoning through the problem that the MC options failed to capture.
\begin{table*}[htbp]
\centering
\footnotesize
\caption{Example of student responses for questions, analyzed by the LLM for incorrect evidence/reasoning}
\label{Example_Student_Response_Codes}
\begin{tabular}{
  p{0.03\linewidth}
  @{\hspace{0.01\linewidth}}
  p{0.48\linewidth}
  @{\hspace{0.01\linewidth}}
  p{0.46\linewidth}}
\toprule
\textbf{Q \#} & \textbf{Student Response} & \textbf{LLM Analysis} \\
\midrule
5 & {\raggedright The velocity in the steel block would be faster than wood if treating both blocks as initially at rest as the momentum of the bullet will not be conserved in the block, however it will be partially transferred into the wood block, causing it to move only slightly frictionless plane whereas with the elastic collision, the the steel will retain at rest or whatever it’s previous velocity was. We use the elastisity principles and knowledge as the problem refers to an elastic (energy and momentum are conserved) and inelastic collision (energy and momentum is not conserved). Thus the wood block would slow the bullet and not retain the momentum whereas the steel would retain it due to elasitc collision. Thus, the bullet bouncing back off further supports the use of this principle.\strut}
& {\raggedright \textbf{Misapplication of Collision Type:} The student correctly claims that the steel block moves faster than the wood block, which aligns with the rubric's correct outcome. However, the reasoning is confused and incorrect. The student mentions "elastic" and "inelastic" collisions but misapplies these concepts. The response incorrectly suggests that momentum is not conserved in the wood block and incorrectly describes the effects of collision types on the blocks' motion.\strut} \\
\midrule
16 & {\raggedright You are able to solve for the final height by invoking the principle of the conservation of linear momentum. The spheres stick together after the collision, indicating that momentum was conserved and kinetic energy was not. That being said, we would have to use concepts from the Energy principle and the Momentum principle to solve for the final height. \strut}
& {\raggedright \textbf{Omission of Energy Conservation:} The student correctly identifies the need to use both the conservation of momentum and energy principles to solve for the final height. However, the reasoning is incomplete as it does not clearly articulate how these principles are applied to find the final height. The mention of momentum conservation is correct, but the explanation lacks detail on how energy conservation is used after the collision to determine the final height.\strut} \\
\midrule
23 & {\raggedright The average horizontal forces exerted by both the rubber ball and the putty ball on the surface is zero. This is due to there being no horizontal forces acting upon them in the scenario given. As there is only a vertical motion. But for the vertical force the rubber ball will exert a greater force than the putty ball. Due to it exerting a larger magnitude of over shorter contact time.\strut}
& {\raggedright \textbf{Misapplication Of Momentum Principles}: The student correctly claims that the rubber ball exerts a greater force than the putty ball, which aligns with the correct conclusion. However, the reasoning is flawed. The student incorrectly states that there are no horizontal forces and focuses on vertical motion without addressing the correct physics principle of impulse or momentum change, which explains why the rubber ball exerts a greater average force due to its rebound. The reasoning does not adequately support the claim with the appropriate physics principles.\strut} \\
\bottomrule
\end{tabular}
\end{table*}
 Table \ref{Example_Student_Response_Codes}  shows a summary of the LLM analysis of the most common misapplied resources by students in this study. The three examples in Table \ref{Example_Student_Response_Codes} come from instances of students making the correct claim (i.e., answer) based on an incorrect explanation (i.e., evidence and reasoning). For instance, looking at the example response for Q05, we found that the student incorrectly explained that momentum was not conserved in the case of the inelastic collision (wood block), but that still led them to the correct claim that the steel block moves faster.  Thus, this student would have likely responded to the MC question correctly even though they were using unproductive conceptual resources. However, the LLM was proven capable of uncovering students' conceptual resources using solely their written responses. Instances such as this demonstrate the full richness of feedback that LLMs can provide to instructors and students at scale, which would help isolate misunderstandings in PER.

\section{CONCLUSIONS, LIMITATIONS, \& FUTURE WORK}
Research-based MC assessments have been used for many years to grade students at scale. However, research \cite{rebello2004effect} has shown that while MC assessments can effectively gauge the percentage of students who can answer the questions correctly, they are unable to uncover the myriad ways in which students explain, in writing, their reasoning about these conceptual physics questions. This study aimed to explore the validity of using LLMs as a possible alternative to gauge students' explanations (evidence and reasoning) in physics, in addition to their answers (claim). 

There are two main conclusions from this work. 1) We have shown that an appropriately prompted LLM (ChatGPT-4o) can effectively assess the overall correctness of students' written responses, including their answer (i.e., claim) and supporting explanations (evidence and reasoning). We find substantial agreement (Cohen's $\kappa = 0.74$) in comparing LLMs with human graders. Therefore, appropriately prompted LLMs can provide a reliable assessment of students' written responses to conceptual physics questions.
2) Further, we find that the LLM-based analysis of the students' written responses to the conceptual physics questions reveals emergent categories of incorrect explanations that are not necessarily covered by the incorrect options on the original MC version of the question.  

Using LLMs to assess students' explanations to conceptual physics questions can potentially facilitate educators to uncover areas of weakness in students' understanding. For instance, we find that LLMs can identify students' reasoning resources that were not included in the normal MC options for the EMCS, and thus, we were able to investigate student resources when working through conceptual questions on energy and momentum. This supports prior literature that LLMs provide a possible source to gauge students' correctness and help uncover students' myriad ways of thinking about physics conceptual questions \citep{weijers2025intuition, latif2025integrating, nlp2024misconceptions}. A key contribution of this study is that we have shown that LLMs can be used not just to grade students' written responses to conceptual questions, but to categorize the myriad conceptual resources that students use to answer these questions.


A key limitation of this study is that we investigated a only subset of questions from the EMCS due to concern for students' fatigue in providing written responses. In future studies, we will expand this study to address students' conceptual resources for other items on the EMCS. Our goal after establishing that the LLM can grade and categorize all students' responses for a well-established conceptual survey will be to see its performance on more complicated physics assessments. However, it seems clear that LLMs offer a promising way to gauge students' misapplied conceptual resources. 
\vspace{-0.25em}

\section{ACKNOWLEDGEMENTS}
This work is supported in part by U.S. National Science Foundation grant 2300645. Opinions expressed are of the authors and not of the Foundation. 
\clearpage

\bibliography{references}
\end{document}